\documentclass[aps,prl,superscriptaddress,twocolumn]{revtex4-2}
\usepackage{bbm}
\usepackage{amssymb}
\usepackage{graphicx}
\usepackage{amsmath}
\usepackage{natbib}
\usepackage{siunitx}
\usepackage{textcomp}
\usepackage{bm}
\usepackage{color}
\usepackage[hidelinks,colorlinks=true,urlcolor=blue,linkcolor=blue,citecolor=blue]{hyperref}

\def\ba{{\boldsymbol a}}
\def\brho{{\boldsymbol \brho}}
\def\bk{{\boldsymbol k}}
\def\bc{{\boldsymbol c}}
\def\bp{{\boldsymbol p}}
\def\bq{{\boldsymbol q}}
\def\br{\boldsymbol{r}}
\def\bj{{\boldsymbol j}}
\def\bv{{\bf v}}
\def\bx{{\bf x}}
\def\bz{{\bf z}}
\def\bG{{\bf G}}
\def\bN{{\bf N}}
\def\bJ{{\boldsymbol J}}
\def\bK{{\boldsymbol K}}
\def\bP{{\bf P}}
\def\la{\langle}
\def\bra{\rangle}
\def\calT{\mathcal{T}}
\def\calM{\mathcal{M}}
\def\calW{\mathcal{W}}
\def\calB{\mathcal{B}}
\def\calH{\mathcal{H}}
\def\calZ{\mathcal{Z}}
\def\calD{\mathcal{D}}
\def\calS{\mathcal{S}}
\def\calA{\mathcal{A}}
\def\calE{\mathcal{E}}
\def\calJ{\mathcal{J}}
\def\p{\hat {\psi}} 
\def\pd{\hat {\psi}^{\dag}}
\def\grad{\mbox{\boldmath $\nabla$}}
\def\Tr{{\brm Tr}}
\def\e{\epsilon}
\def\ve{\varepsilon}
\def\vphi{a}
\def\pa{\partial}
\def\nn{\nonumber}
\def\t{\tau}
\def\kbar{\bar {k}}
\def\brbar{\bar {r}}
\def\nbar{\bar {n}}
\def\la{\langle}
\def\ra{\rangle}
\begin{document}

\title{Strongly interacting Bose-Fermi mixture: mediated interaction, phase diagram and sound propagation}
\author{Xin Shen}
\affiliation{College of Sciences, China Jiliang University, Hangzhou 310018, China}
\author{Nir Davidson}
\affiliation{Department of Physics of Complex Systems, Weizmann Institute of Science, Rehovot 7610001, Israel}
\author{Georg M. Bruun}
\affiliation{Center for Complex Quantum Systems, Department of Physics and Astronomy,
Aarhus University, Ny Munkegade, DK-8000 Aarhus C, Denmark}
\author{Mingyuan Sun}
\email{mingyuansun@bupt.edu.cn}
\affiliation{State Key Lab of Information Photonics and Optical Communications, Beijing University of Posts and Telecommunications, Beijing 100876, China}
\affiliation{School of Science, Beijing University of Posts and Telecommunications, Beijing 100876, China}
\author{Zhigang Wu}
\email{wuzg@sustech.edu.cn}
\affiliation{Shenzhen Institute for Quantum Science and Engineering, Southern University of Science and Technology, Shenzhen 518055, China.}

\date{\today }
\begin{abstract}
 Motivated by recent surprising experimental findings, we develop a strong-coupling theory for Bose-Fermi mixtures capable of treating resonant 
 inter-species interactions while satisfying the compressibility sum rule. We show that the mixture can
 be stable at large interaction strengths close to resonance, in agreement with the experiment but at odds with the widely used 
 perturbation theory. We also calculate the sound velocity of the Bose gas in the $^{133}$Cs-$^6$Li mixture, again finding good agreement with the experimental observations both at weak and strong interactions. A central ingredient of our theory is the generalization of a fermion mediated interaction to strong Bose-Fermi scatterings and to finite frequencies.
This further leads to a predicted hybridization of the sound modes of the Bose and Fermi gases, which can be directly observed using Bragg spectroscopy.

\end{abstract}
\maketitle
{\it Introduction.}---The interest in mixtures of bosonic and fermionic quantum fluids has long predated the discovery of ultracold atomic gases. Indeed, as early as in the 1960s  ${^3}$He-${^4}$He solutions were studied by H. London and others~\cite{London1962}, which led to the creation of an indispensable workhorse of low temperature experiments---the dilution refrigerator~\cite{HALL1966}. For ultracold atomic gases, the Bose-Fermi mixture is not only practically valuable for sympathetically cooling the Fermi gas~\cite{Schreck2001a,Schreck2001b}, but also serves as a versatile platform for studying a variety of physics, including polarons~\cite{Massignan2014,Scazza2022}, mediated interactions~\cite{2008Santamore,Nishida2010,Yu2012,Kinnunen2015,Suchet2017,
Camacho-Guardian2018,DeSalvo2019,Edri2020}, unconventional pairing~\cite{Wu2016,Kinnunen2018} and dual superfluidity~\cite{Barbut2014,Yao2016,Roy2017}. Due to its importance, more than a dozen  different Bose-Fermi mixtures have so far been realized and studied experimentally (see Ref.~\cite{Onofrio2016} for a review).

Since the inter-species interaction can be tuned in an atomic Bose-Fermi mixture, the first fundamental question concerns its stability and 
miscibility~\cite{Klaus1998,Viverit2000,Viverit2002,Modugno2002,Albus2002,Capuzzi2003,chui2004,Salasnich2007,
Marchetti2008,Yu2011,Lous2018,Kim2018,Manabe2021}. For a  weakly interacting Bose-Einstein condensate (BEC) mixed with a single-component Fermi gas, perturbation theory predicts that a sufficiently large Bose-Fermi scattering length will lead to the collapse of the system on the attractive side~\cite{Viverit2000,Modugno2002,chui2004} and to phase separation on the repulsive side~\cite{Capuzzi2003,Marchetti2008,Lous2018,Kim2018}. At typical atomic gas densities, the predicted critical values of the scattering length are quite small such that  perturbation theory is expected to be valid. The recent experimental results 
for the $^{133}$Cs-$^6$Li mixture have therefore come as a surprise~\cite{Patel2022}. By measuring the bosonic sound propagation at varying Bose-Fermi scattering lengths, the experiments found that the mixture regains its stability near the inter-species Feshbach resonance, in contradiction with the perturbation theory~\cite{Patel2022}. 

In order to understand this puzzling phenomenon and more broadly the properties of resonant Bose-Fermi mixtures, we develop a strong-coupling approach based on the many-body Bose-Fermi scattering matrix. Importantly, our theory satisfies the compressibility sum rule~\cite{Watabe2014,Pitaevskii2016}, which plays a crucial role in determining the stability of the mixture. With this approach, we first obtain the zero-temperature phase diagram of the mixture corresponding to the experimental setup. 
The predicted region of stability is consistent with the experimental observation but differs significantly from that of the perturbative theory near resonance.  
An integral part of our theory is a generalization of the well-known Ruderman–Kittel–Kasuya–Yosida (RKKY) fermion mediated interaction~\cite{RK1954} to the regime of strong Bose-Fermi scattering. Based on this  interaction, we further calculate the speed of sound in the BEC and find reasonable agreement with the recent experiment for all interaction strengths. 
Lastly, we show that the retarded nature of this mediated interaction leads to an intriguing hybridization of the BEC sound mode and an induced fermionic zero sound mode, which can be observed by Bragg spectroscopy.

\begin{figure}[tbp]
	\centering
	\includegraphics[width=8.3cm]{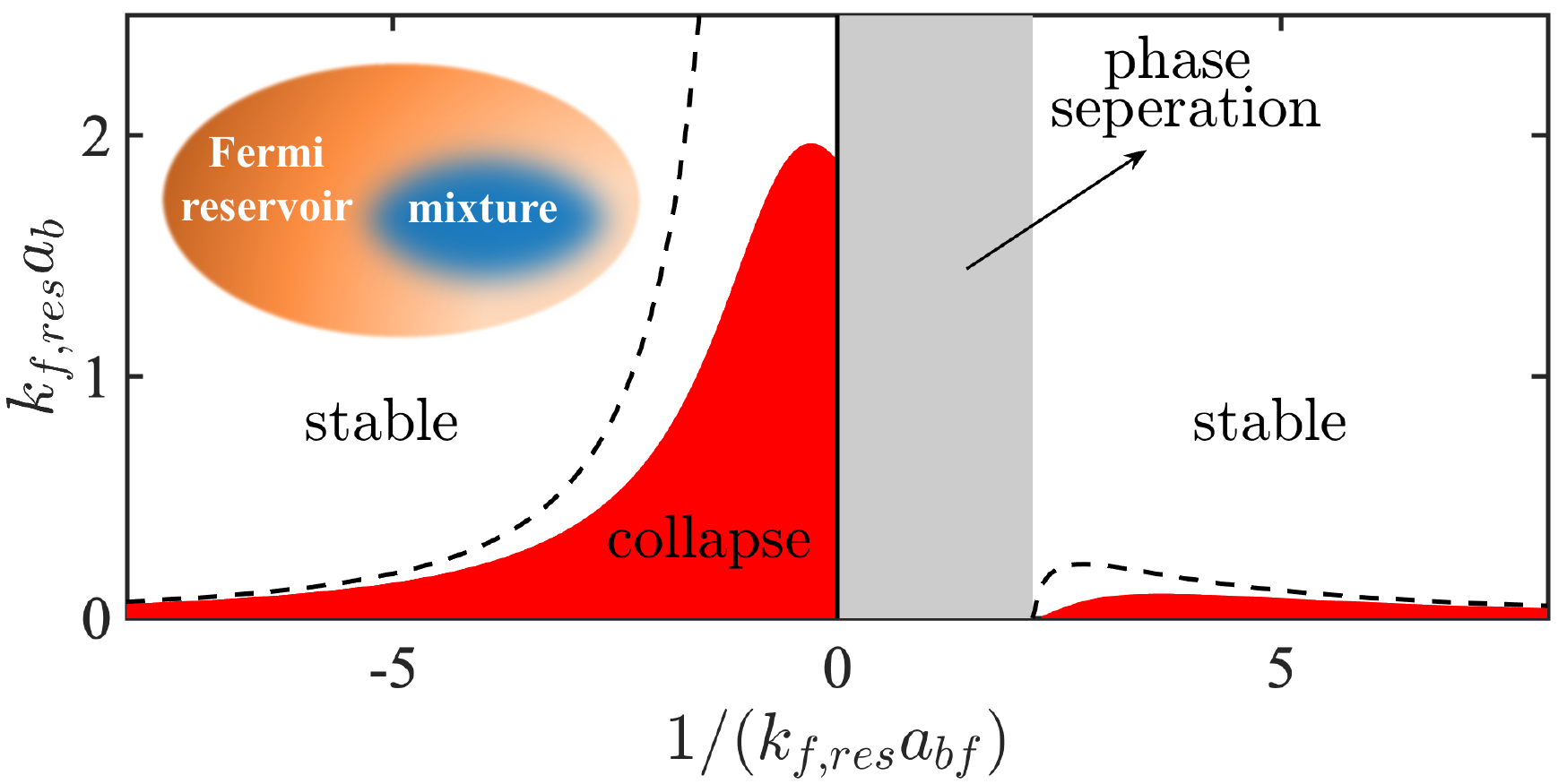}
	\caption{Phase diagram of the $^{133}$Cs-$^{6}$Li mixture with density ratio $n_b/n_{f,res} = 10$, where $n_{f,res}$ is the reservoir fermion density and $k_{f,res} \equiv (6\pi^2n_{f,res})^{1/3}$ is the corresponding Fermi momentum. The dashed lines are the stability boundaries calculated by the perturbation theory.}
	\label{phase}
\end{figure}

{\it Bose-Fermi mixture.}---We consider a mixture of a weakly interacting  BEC of bosons with mass $m_b$ and a non-interacting gas of fermions with mass $m_f$ at zero temperature and in a configuration that is representative of many current experimental systems~\cite{DeSalvo2019,Edri2020,Patel2022}. Namely, the BEC is completely immersed in a spatially much larger Fermi gas such that the Fermi gas surrounding the bosons acts effectively as a reservoir for the Fermi gas inside the mixture; this is illustrated in the inset of Fig.~\ref{phase}. The Hamiltonian for the mixture is 
\begin{align}
\hat H =&  \sum_{\bp \neq 0} \big [( \epsilon_{b,\bp} +2g_b n_b) \hat b_\bp^\dag \hat b_\bp + \frac{1}{2}g_b n_b( \hat b_\bp^\dag \hat b_{-\bp}^\dag  + h.c.)\big ]\nonumber\\
+&\sum_\bp  \epsilon_{f,\bp}\hat f_\bp^\dag \hat f_\bp+g_{bf} \sum_{\bp\bp'\bq}\hat f^\dag_\bp \hat b^\dag_{\bp'} \hat b_{\bp'+\bq}\hat f_{\bp-\bq}, 
\label{Hamiltonian}
\end{align}
where $ \hat b_\bp^\dag (\hat f_\bp^\dag)$ creates a boson (fermion) of momentum $\bp$ and energy $\epsilon_{i,\bp}=\bp^2/2m_i$ with $i=b (f)$. We have 
used the Bogoliubov theory to describe the BEC with density 
$n_b$ and  interaction strength $g_b = {4\pi a_b}/{m_b}$, where $a_b$ is the bosonic scattering length. Similarly,  the Bose-Fermi interaction strength $g_{bf}$ is determined by the corresponding scattering length $a_{bf}$. Here we use units where $\hbar$ and the system volume are unity.

\begin{figure}[tbp]
	\includegraphics[width=8.4cm]{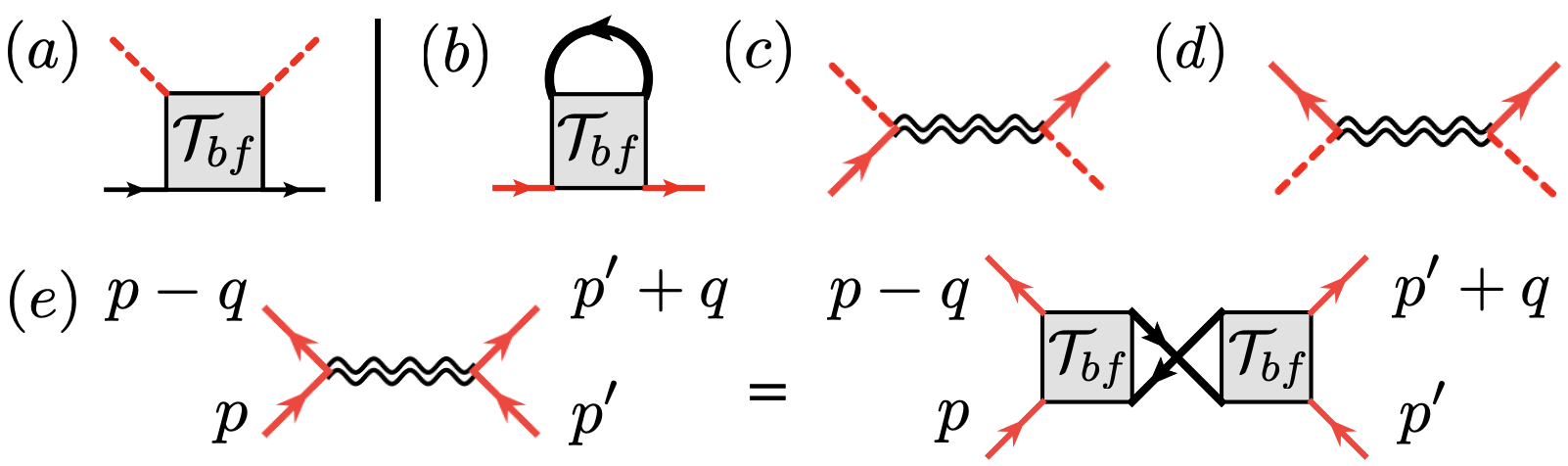}
	\caption{Self-energy diagrams for fermions (a) and bosons (b)-(d) due to the Bose-Fermi interaction. The (dashed) red line denotes a (condensate) 
	boson and the black line a fermion. {The wavy line is the fermion mediated interaction in Eq.~\eqref{Gamma}.} } 
	\label{se}
\end{figure}

{\it Strong-coupling theory.}--- In order to describe  strong Bose-Fermi interactions, we use a Green's function approach 
with the Bose-Fermi scattering matrix as a basic building block~\cite{Fratini2010,Fratini2012,Guidini2014,Guidini2015}. 
Within this framework, the fermionic Green's function is given by (see Fig.~\ref{se}(a)),
\begin{align}
G_f (p) = \frac1{i\omega_p - (\epsilon_{f,\bp}- \mu_f)- n_b \calT_{bf}(p)},
\label{Gf}
\end{align}
where $\mu_f$ is the chemical potential of the fermions inside the mixture,  $\omega_p$ is the Matsubara frequency and $p \equiv (i\omega_p,\bp)$. The scattering matrix between a boson and a Fermi gas of density $n_f$ is 
\begin{align}
\calT_{bf}(p) = \frac{g_{bf}}{1-g_{bf}\Pi_{bf}(p)}
\label{Tmatrix}
\end{align}
with the pair propagator
\begin{equation}
\Pi_{bf}(p) = \int\! \frac{d^3\bk}{(2\pi)^3} \left [ \frac{1-n_{\rm FD}(\bk)}{i\omega_p -\epsilon_{b,\bp-\bk}-{\xi_{f,\bk}} } + \frac{2m_r}{k^2}\right ].\label{Pairprop}
\end{equation}
Here $\xi_{f,\bk} = \bk^2/2m_f - k_f^2/2m_f$ with $k_f \equiv (6\pi^2 n_f)^{1/3}$ and  $n_{\rm FD}(\bk)= \theta(\xi_{f,\bk})$ is the Fermi-Dirac distribution. We have neglected the effects of the BEC on the pair propagator, 
which is a good approximation for $n_ba_b^3\ll 1$. The last term in Eq.~\eqref{Pairprop} regularizes the divergence coming from the momentum independence of $g_{bf}$ so that one can establish the relation $g_{bf}=2\pi a_{bf}/m_r$, where $m_r=m_bm_f/(m_b+m_b)$ is  the reduced mass.

 To the lowest order in the scattering matrix, the effects of the Fermi gas on the BEC are captured by the self-energy diagram shown in Fig.~\ref{se}(b). 
As shown in the Sup.\ Mat.~\cite{SM}, in order to fulfill the compressibility sum rule one also needs to include the diagrams in 
Fig.~\ref{se}(c)-(d), which are second order in $\mathcal T_{bf}$. Incorporating also the usual Bogoliubov self-energies due to the weak 
boson-boson scattering, we find  
\begin{align}
  \Sigma_{11}(p)& =2n_b g_b+\sum_k G_f(k)\calT_{bf}(k+p)+n_b \Gamma_{\rm mi}(p,0;p)\nonumber\\
  \Sigma_{12}(p)& = n_b g_b +n_b \Gamma_{\rm mi} (p,-p;p) 
  \label{S12}
\end{align}
as the normal and anomalous self-energies of the BEC. Here we have defined the generalized fermion mediated interaction (shown in Fig.~\ref{se}(e))
\begin{align}
\Gamma_{\rm mi}(p,p';q) = \sum_k &G_f (k)G_f(k+q) \nn \\
&\times \calT_{bf}(p+k) \calT_{bf}(p'+k+q), 
\label{Gamma}
\end{align}
where  $\sum_k \equiv T\sum_{i\omega_k}\int\frac{d^3\bk}{(2\pi)^3}$ with temperature $T$. The normal and anomalous Green's functions of the BEC can be obtained from the coupled equations~\cite{AGD}
\begin{align}
\label{BG1}
\left [{G^{0}}(p)^{-1} -\Sigma_{11}(p) \right ]G_{11}(p) - \Sigma_{12}(p) G_{12}(p) &= 1;  \\
\left [{G^{0}}(-p)^{-1} -\Sigma_{11}(-p) \right ]G_{12}(p) - \Sigma_{12}(p) G_{11}(p) &= 0,
\label{BG2}
\end{align}
where $G^{0}(p) = (i\omega_p -\epsilon_{b,\bp}+\mu_b)^{-1}$ and $\mu_b$ is the bosonic chemical potential.  To ensure that the bosonic spectrum is gapless, the chemical potential  must satisfy the Hugenholtz-Pines theorem $\mu_b = \Sigma_{11}(0)- \Sigma_{12}(0)$~\cite{AGD}. From Eq.~(\ref{S12}) we find
\begin{align}
\mu_b = n_b g_{b} +\sum_k G_f(k)\calT_{bf}(k).
\label{mub}
\end{align} 
Solving Eqs.~(\ref{BG1})-(\ref{mub}) yields  the bosonic Green's functions, which  can be used to calculate the Bogoliubov spectrum $E_{b,\bp}$ and other physical properties.

{\it Phase diagram.}---We first use our strong-coupling theory to construct the zero temperature phase diagram spanned by the two scattering lengths
$a_b$ and $a_{bf}$. The stability and miscibility of the mixture are determined by two conditions~\footnote{If the Bose gas is not confined by a trapping potential such that the mixture is completely bounded by the pressure of the reservoir, a third condition of equal pressure is required for the discussion of miscibility. }: (a) the chemical potential $\mu_f$ of the Fermi gas within the mixture equals  that of the reservoir $\mu_{f,res}$;  (b) the compressibility of the BEC under a fixed fermion chemical potential is positive definite~\cite{Bardeen1967}, i.e., $\left.{(\pa \mu_b}/{\pa n_b})\right|_{\mu_{f}} \geq 0$. The first condition places a constraint on the fermion density inside the mixture while the second ensures that the mixture is stable against collapse. In Fig.~\ref{phase}, we show the phase diagram obtained from 
these conditions for the experimentally relevant case of a  $^{133}$Cs-$^{6}$Li mixture with density ratio $n_b/n_{f,res} = 10$. In the following we discuss in detail how this phase diagram is obtained. 

\begin{figure}[b]
	\centering
	\includegraphics[width=8.3cm]{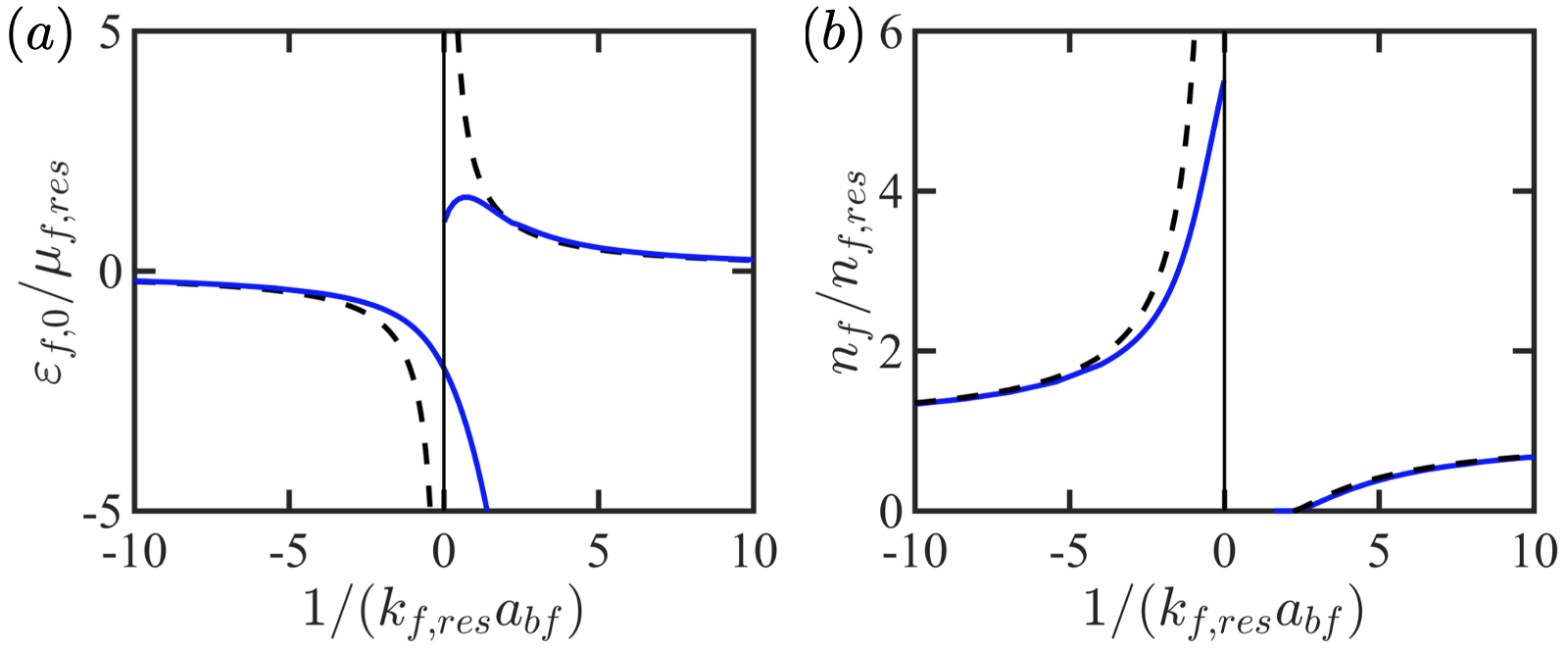}
	\caption{(a) The attractive and repulsive fermion quasi-particle branch $\varepsilon_{f,\bp=0}$ for the $^{133}$Cs-$^{6}$Li mixture of Fig.~\ref{phase}. (b) Corresponding fermion densities inside the mixture. Solid and dashed lines are the strong-coupling and perturbation theory respectively.  We assume for the moment that the mixture is always stable. } 
	\label{fermion_nf}
\end{figure}

Using the condition $\mu_f = \mu_{f,res}$, the fermionic quasi-particle dispersion $\varepsilon_{f,\bp}$ is determined from the poles of $G_{f}(p)$ and the fermion density inside the mixture is calculated as $n_f  =  \sum_p G_f(p)$.
We find that similar to the  so-called Bose polaron, i.e., a single fermion in a BEC~\cite{Jorgensen2016,Hu2016,Yan2020}, the 
fermion Green's function also has two quasi-particle branches: an attractive and a repulsive one; the attractive (repulsive) branch has negative (positive) energy and takes most of the spectral weight for $a_{bf}<0$ $ (a_{bf}>0)$. These two branches are shown  in Fig.~\ref{fermion_nf}(a) 
for the $^{133}$Cs-$^{6}$Li mixture.  Hence, we assume that the fermions occupy the attractive branch for $a_{bf}<0$ and the repulsive branch for $a_{bf}>0$. 
Since $\mu_f$ is fixed by the reservoir, it follows that the density $n_f$ of fermions occupying the attractive branch increases  as $a_{bf}$ is tuned from a small negative value to resonance; the opposite is true for fermions occupying the repulsive branch. This is shown in Fig.~\ref{fermion_nf}(b).  Thus, in the latter case the fermion density inside the mixture vanishes beyond a critical value of $a_{bf}$, leading to phase separation between the fermions and the bosons indicated by the grey region in Fig.~\ref{phase}.
 \begin{figure}[tbp]
	\centering
	\includegraphics[width=9cm]{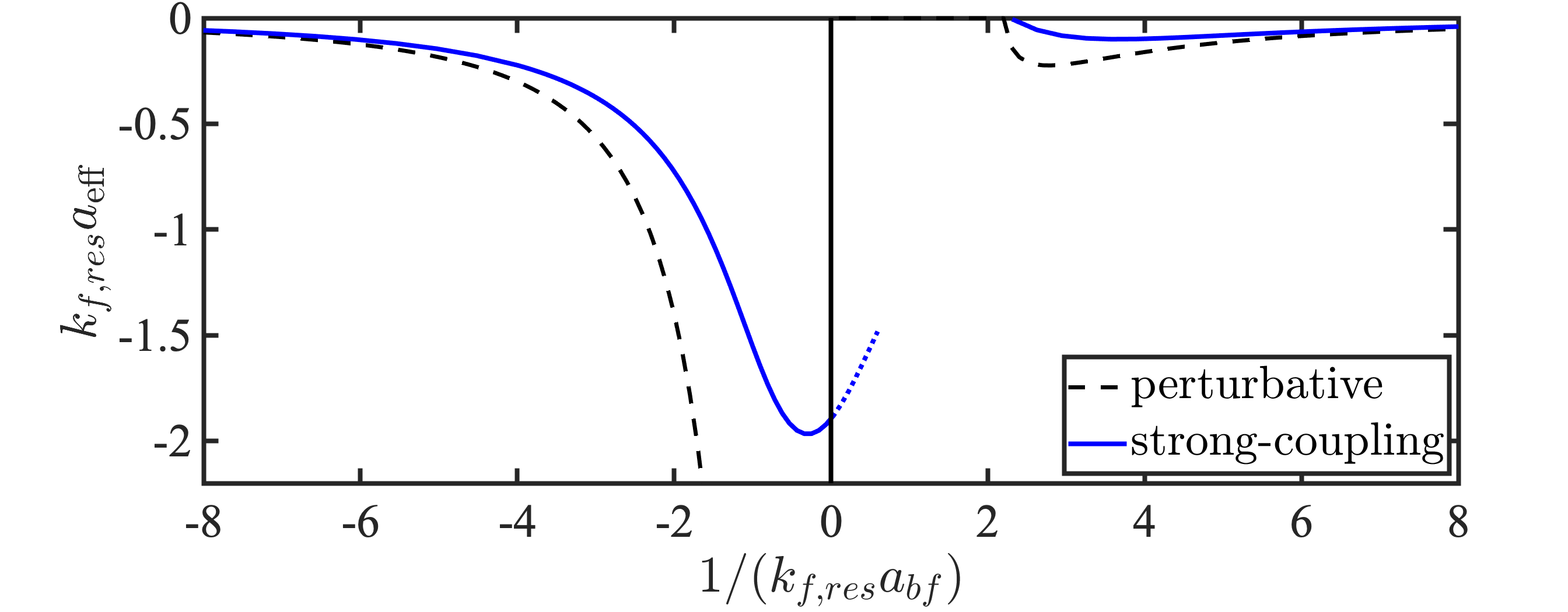}
	\caption{The effective scattering lengths from the mediated interaction, calculated by the perturbation theory  and the strong-coupling theory. The blue dotted line shows the behavior of $a_{\rm eff}$ assuming the fermions stay on the attractive branch beyond the resonance. The results are for the $^{133}$Cs-$^{6}$Li mixture of Fig.~\ref{phase}. } 
	\label{mi}
\end{figure}

Outside the region of phase separation, the stability of the mixture is determined by the compressibility of the BEC. From  Eq.~(\ref{mub}), we find
\begin{align}
\left.\frac{\pa \mu_b}{\pa n_b} \right |_{\mu_{f}}  = \frac{4\pi }{m_b}\left [ a_b + \frac{m_b}{4\pi }\Gamma_{\rm mi}(0,0;0)\right ].
\label{comp}
\end{align}
This relation naturally leads to an  effective scattering length from the fermion mediated interaction given by  
\begin{align}
a_{\rm eff} \equiv \frac{m_b}{4\pi}\Gamma_{\rm mi}(0,0;0).
\end{align}
It then follows from Eq.~\eqref{comp} that the BEC collapses when the total scattering length $a_b+a_{\rm eff}$ turns negative.

 In the weak Bose-Fermi interaction limit, we can replace $\calT_{bf} $ by $ g_{bf}$ and the fermion mediated interaction in Eq.~(\ref{Gamma}) reduces 
  to the familiar RKKY form  $\Gamma_{\rm mi}(q) = g_{bf}^2\chi^{(0)}_f(i\omega_q,\bq) $~\cite{Yu2012,Kinnunen2015},
  where $\chi^{(0)}_f(i\omega_q,\bq)$ is the Lindhard function of a free Fermi gas. Since $\chi_f^{(0)}(0,\bq)  = -(m_fk_f/2\pi^2)(1-q^2/8k_f^2)$ in the long wavelength limit, 
  second order perturbation theory predicts that $a_{\rm eff} =-({1}/{2\pi})({m_f}/{m_b}+m_b/m_f+ 2)k_f a_{bf}^2 $~\cite{Patel2022,SM}. 
In Fig.~\ref{mi}(c) we compare this result against that calculated by our strong-coupling theory. We find that while the two approaches agree for weak coupling as 
expected, the strong coupling result for  $a_{\rm eff}$ is significantly smaller close to unitarity. 

This has important consequences for the phase diagram. Since the BEC collapses for $a_b+a_{\rm eff}<0$ as discussed above, the values of $a_{\rm eff}$ shown in Fig.~\ref{mi} 
 directly give the boundaries for the collapse regions shown in 
 Fig.~\ref{phase}.  While perturbation theory predicts a collapse region that extends to arbitrarily large values of $a_b$ as unitarity $1/a_{bf}=0$ is 
 approached, our  strong-coupling theory predicts a much smaller collapse region bounded by a maximum value of $a_b$ near resonance. 
 It follows that  the mixture is  stable even at resonance provided that $a_b$ is sufficiently large. In Fig.~\ref{phase}, we see  that the 
 region of stability of the Bose-Fermi mixture indeed is significantly larger than predicted from perturbation theory.

{\it Bosonic sound propagation.}---We next turn to the discussion of bosonic sound propagation observed recently in a strongly-interacting $^{133}$Cs-$^{6}$Li  mixture~\cite{Patel2022}. As usual,  the Bogoliubov sound velocity in the BEC is defined from the Bogoliubov spectrum as $c_b=\lim_{\bp\rightarrow 0 }E_{b,\bp}/|\bp|$. In a pure BEC, this velocity is given by $c_b^{(0)} = \sqrt{n_bg_b/m_b}$ which 
coincides with that defined by the compressibility $c_{b,\rm com} = \sqrt{({n_b}/{m_b}){\pa \mu_b}/{\pa n_b}}$~\cite{Pitaevskii2016}.
Interestingly, these two quantities are not equal in the Bose-Fermi mixture due to the retarded nature of the fermion mediated interaction. 
Retardation effects can however be ignored when the Fermi velocity $v_f$ is much larger than the sound velocity in the pure BEC, i.e., when
$c_b^{(0)}/v_f=(m_f/m_b)\sqrt{(2/3\pi)(n_b/n_f)(k_f a_b)} \ll 1$. This is indeed the  
case for the $^{133}$Cs-$^{6}$Li mixture in Ref.~\cite{Patel2022} due to the very small Fermi-Bose mass ratio.

 \begin{figure}[tbh]
	\centering
	\includegraphics[width=8.2cm]{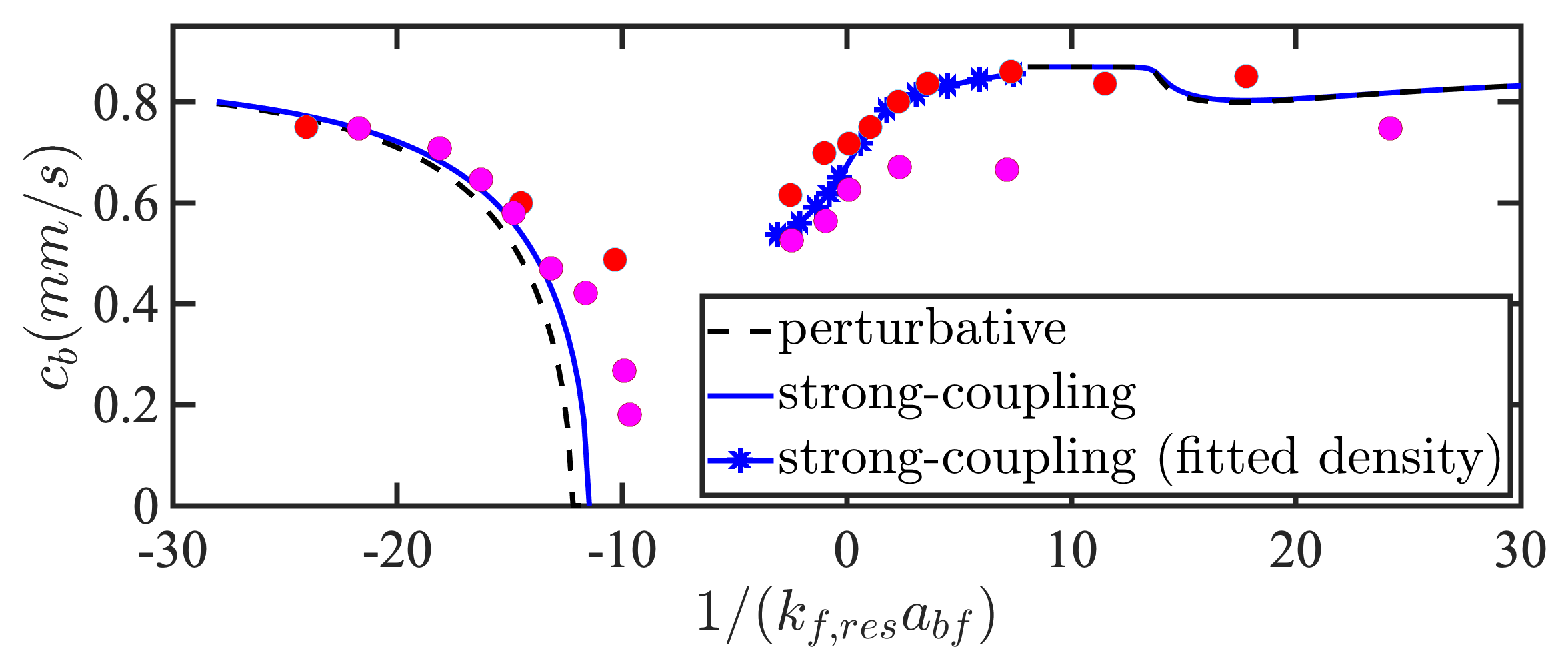}
	\caption{Comparison of our strong-coupling theory for the BEC sound velocity 
	to  the experimental results (dots) in Ref.~\cite{Patel2022}.}
	\label{sound}
\end{figure}
We therefore use the compressibility formula to calculate the sound velocity and compare the results with the recent experiment.
In order to do this, we must first analyze the experimental procedure. 
In Ref.~\cite{Patel2022}, the mixture is first prepared at a small value of $a_{bf}$ on either side of the Feshbach resonance and is subsequently ramped
 to a target value of $a_{bf}$ within a fixed duration of time. This process is 
 approximately adiabatic for  small target values $a_{bf}$, but highly non-adiabatic for target values  in the resonant regime. 
 Consequently, a significant fraction of the fermions will not remain on the same quasi-particle branch under such 
 non-adiabatic ramps  due to the Landau-Zener transitions~\cite{Kohler2006,Bortolotti2008}. Furthermore, heavy losses of atoms are observed in experiments near 
  resonance~\cite{Patel2022}. For these reasons, one must expect that near resonance the experimental values for the fermion densities inside the mixture will be much smaller than those predicted by our thermal equilibrium theory described above. 
  To make comparisons with experiments conducted near resonant $a_{bf}$, we therefore treat $n_f$ as a fitting parameter using 
  $n_f/n_{f,res}= 0.006 +0.05\times(k_fa_{bf})^{-1/2} $ for $a_{bf}>0$ and $n_f/n_{f,res}= 0.006 -0.0002\times(k_fa_{bf})^{-3} $ for $a_{bf}<0$ as 
   suggested by the behavior of the loss data in experiments~\cite{Patel2022}. Other parameters are the same as those in the experiment, i.e.,  $n_b \approx 1.87\times 10^{19} m^{-3} $, $n_{f,res} \approx 3 \times 10^{17} m^{-3}$ and $a_b = 270 a_0$ where $a_0$ is the Bohr radius.
   As can be seen in Fig.~\ref{sound}, for small $a_{bf}$ both perturbative and strong-coupling theory with no fitting of $n_f$ agree well with experiments although the latter performs slightly better. For resonant $a_{bf}$, however, perturbative theory predicts no sound propagation while the strong-coupling theory with fitted $n_f$ reproduces the experimental measurements well. 

{\it Retardation and induced fermionic zero sound.}---The Fermi-Bose mass ratio is much larger for a $^{23}$Na-$^{40}$K mixture~\cite{Wu2012,Park2012,Zhu2017} compared to a $^{133}$Cs-$^{6}$Li mixture, and it follows from the arguments given above that retardation effects must  be significant for the former. A remarkable consequence of this is the possibility of exciting an induced fermionic zero sound mode through a bosonic density perturbation. It is known that in a Bose-Fermi mixture the non-interacting fermions can also experience a mediated interaction due to the Bose gas, which can lead to a fermionic zero sound mode with a speed  $\sim v_f$~\cite{Yip2001,Capuzzi2001,Santamore2004}. When the Bose-Fermi interaction is strong and the zero sound velocity is comparable to that in the pure BEC, 
we anticipate a strong coupling of these two modes. 

 \begin{figure}[tb]
	\centering
			\includegraphics[width=8.3cm]{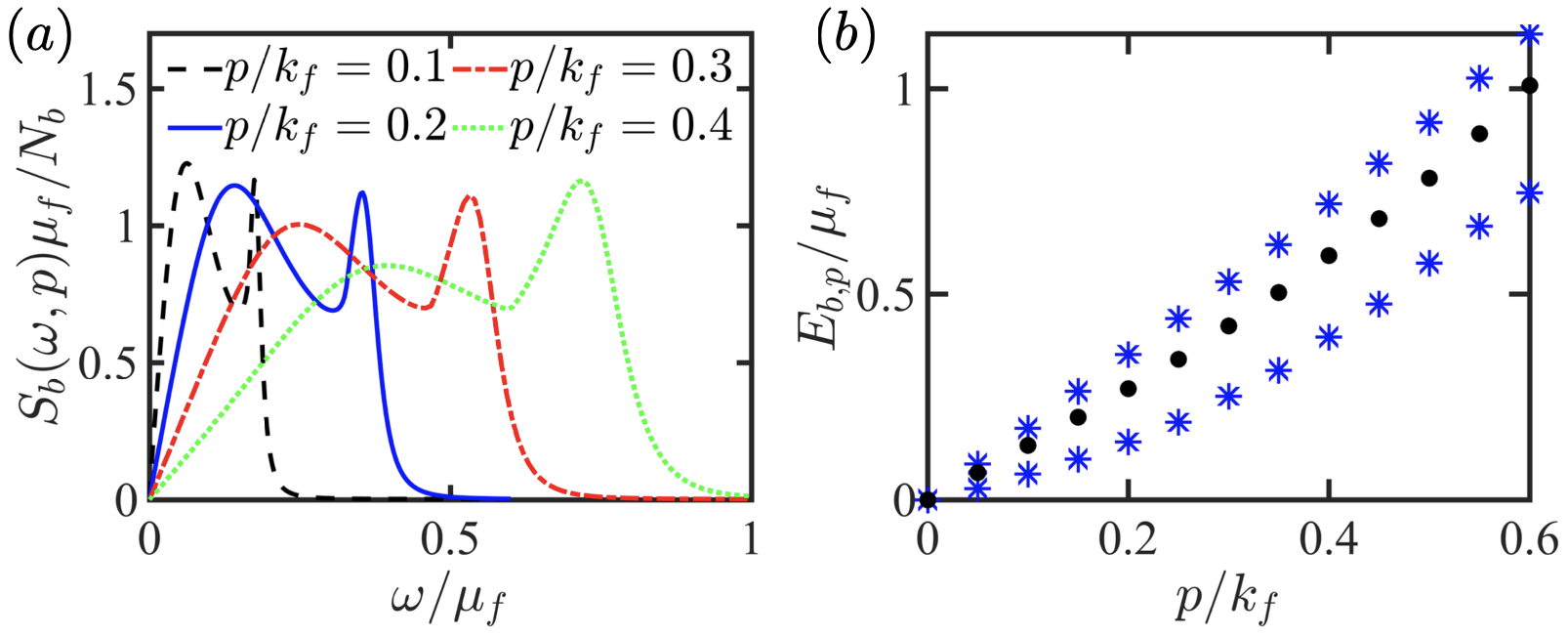}
	\caption{(a) Dynamic structure factor $S_b(\omega,\bq)$ of the BEC at different momenta for a strongly-interacting $^{23}$Na-$^{40}$K mixture. (b) Excitation spectrum obtained from the peaks of $S_b(\omega,\bq)$  
	(blue asterisks) and for a pure BEC (black dots).} 
	\label{spectrum}
\end{figure}

In order to demonstrate this, we turn to the calculation of the dynamic structure factor of the BEC, which also gives the sound spectrum~\cite{Pitaevskii2016} and can be directly probed by Bragg spectroscopy~\cite{Ozeri2005,Li2022}. It is defined as
 \begin{align}
 S_b(\omega,\bp) \equiv \frac{1}{\pi} {\rm Im}\chi_b(i\omega_p\rightarrow \omega + i0^+,\bp).
 \end{align}
 Here $\chi_b(p)$ is the density-density  response function of the BEC and is  given by 
 $\chi_b(p) = -2N_b[G_{11}(p)+G_{11}(-p)+2G_{12}(p)]$ within the Bogoliubov framework, where $N_b$ is the total number of bosons. We now calculate $S_b(\omega,\bq)$ for a $^{23}$Na-$^{40}$K mixture with 
 $n_b/n_f = 10$, $k_fa_b = 0.067$ and $1/(k_f a_{bf}) = -3$, which   yields $c_b^{(0)}/v_f \sim 0.65$. 
 As shown in Fig.~\ref{spectrum}(a), $S_b(\omega,\bq)$  exhibits a double peak structure  indicating the presence of two modes,
   in stark contrast with the single peak structure at small $a_{bf}$ or for $c^{(0)}_b/v_f\ll 1$~\cite{SM}. Figure \ref{spectrum}(b) plots the dispersion 
   of these two modes, which are compared to the single mode in a pure BEC. This explicitly  demonstrates that a fermionic zero sound mode indeed can hybridize with the Bogoliubov sound mode and manifest itself in the excitation spectrum of the BEC.

{\it Concluding remarks.}--- 
We have developed a strong-coupling theory for  the ground state and collective excitations of strongly interacting Bose-Fermi mixtures, emphasizing the role of a generalized mediated interaction.  Our theory agrees well with recent experimental results for  a resonant $^{133}$Cs-$^6$Li mixture, which the much used perturbation theory fails to account for. Furthermore, we show that new, interesting physics caused by retardation of the generalized mediated interaction can be revealed by the bosonic dynamic structure factor and observed in future experiments. Finally, in light of the many different mixtures  being studied experimentally, our approach may be used to systematically explore the effects of mass and density ratio on properties of strongly interacting Bose-Fermi mixtures.

\textit{Acknowledgement}. We thank Shizhong Zhang and  Ren Zhang for helpful discussions. This work is supported by National Key R$\&$D Program of China (Grant No. 2022YFA1404103), NSFC (Grant No.~11974161), NSFC (Grant No.~12004049), NSFC (Grant No.~12104430), Shenzhen Science and Technology Program (Grant No.~KQTD20200820113010023) and Key-Area Research and Development Program of Guangdong Province (Grant No.~2019B030330001).

%


	\onecolumngrid
	\clearpage
	\begin{center}
		\textbf{\large Supplemental Material for ``Strongly interacting Bose-Fermi mixture: mediated interaction, phase diagram and sound propagation"}
	\end{center}
	
\def\ba{{\boldsymbol a}}
\def\brho{{\boldsymbol \brho}}
\def\bk{{\boldsymbol k}}
\def\bc{{\boldsymbol c}}
\def\bp{{\boldsymbol p}}
\def\bq{{\boldsymbol q}}
\def\br{\boldsymbol{r}}
\def\bj{{\boldsymbol j}}
\def\bv{{\bf v}}
\def\bx{{\bf x}}
\def\bz{{\bf z}}
\def\bG{{\bf G}}
\def\bN{{\bf N}}
\def\bJ{{\boldsymbol J}}
\def\bK{{\boldsymbol K}}
\def\bP{{\bf P}}
\def\la{\langle}
\def\bra{\rangle}
\def\calT{\mathcal{T}}
\def\calM{\mathcal{M}}
\def\calW{\mathcal{W}}
\def\calB{\mathcal{B}}
\def\calH{\mathcal{H}}
\def\calZ{\mathcal{Z}}
\def\calD{\mathcal{D}}
\def\calS{\mathcal{S}}
\def\calA{\mathcal{A}}
\def\calE{\mathcal{E}}
\def\calJ{\mathcal{J}}
\def\p{\hat {\psi}} 
\def\pd{\hat {\psi}^{\dag}}
\def\grad{\mbox{\boldmath $\nabla$}}
\def\Tr{{\brm Tr}}
\def\e{\epsilon}
\def\ve{\varepsilon}
\def\vphi{a}
\def\pa{\partial}
\def\nn{\nonumber}
\def\t{\tau}
\def\kbar{\bar {k}}
\def\brbar{\bar {r}}
\def\nbar{\bar {n}}
\def\la{\langle}
\def\ra{\rangle}
\setcounter{section}{0}  
\renewcommand{\theequation}{S\arabic{equation}}
\setcounter{equation}{0}  
\renewcommand{\thefigure}{S\arabic{figure}}
\setcounter{figure}{0}  
\renewcommand{\thetable}{S\arabic{table}}
\setcounter{table}{0}  

This supplemental material contains the following five sections: (I) perturbation theory (II) compressibility sum rule (III)  fermion quasi-particle dispersion and density (IV)  effective scattering length and (V) bosonic dynamic structure factor.

\section{Perturbation theory}
\label{pt}
In this section, we show that our strong-coupling theory recovers the well-known perturbation theory~\cite{Viverit2000SM,Viverit2002SM} in the limit that $g_{bf} \rightarrow 0$. In this weak interaction limit  we have $\calT_{bf} \rightarrow g_{bf}$. From Eq.~(2) of the main text we find that the fermion Green's function is given by 
\begin{align}
G_f (i\omega_p,\bp) = \frac{1} {i\omega_p - (\epsilon_{f,\bp}-\mu_f)- n_b g_{bf} },
\label{Gf0}
\end{align}
where $\epsilon_{f,\bp} = \bp^2/2m_f$. Under the condition that $\mu_f = \mu_{f,res}$, the fermion quasi-particle dispersion is 
\begin{align}
\varepsilon_{f,\bp} &= \epsilon_{f,\bp}  + n_b g_{bf} \nn \\& = \mu_{f,res} \left  [ \left(\frac{p}{k_{f,res}}\right)^2+ \frac{2}{3\pi}\frac{n_b}{n_{f,res}}\left ( 1 + \frac{m_f}{m_b} \right )  k_{f,res}a_{bf} \right ],
\label{fdmf}
\end{align}
where $k_{f,res} = \sqrt{2m_f\mu_{f,res}}$ is the reservoir Fermi momentum  and $n_{f,res} = k_{f,res}^3/6\pi^2$ is the reservoir density. 
The fermion density is then given by
\begin{align}
n_f = \frac{k_f^3}{6\pi^2},
\end{align}
where the Fermi momentum for the Fermi gas in the mixture is 
\begin{align}
k_f = \sqrt{2m_f(\mu_{f,res}-n_bg_{bf})} = k_{f,res} \sqrt{1-\frac{2}{3\pi}\frac{n_b}{n_{f,res}}\left ( 1 + \frac{m_f}{m_b} \right )  k_{f,res}a_{bf}  }.
\label{kf0}
\end{align}
The perturbation results in Fig.~3 of the main text are obtained from Eqs.~(\ref{fdmf})-(\ref{kf0}).

\begin{figure}[th]
	\centering
	\includegraphics[width=12cm]{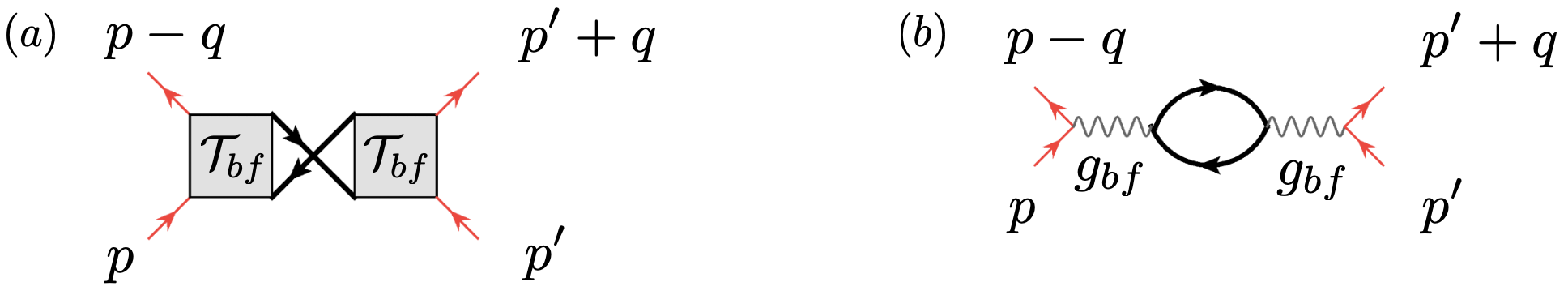}
	\caption{(a) Feynman diagram for the mediated interaction within the strong-coupling theory (b) Feynman diagram for the mediated interaction within the perturbation theory. }
	\label{Pif}
\end{figure}
Substituting Eq.~(\ref{Gf0}) in Eq.~(6) of the main text and setting $\calT_{bf} = g_{bf}$ (see Fig.~\ref{Pif}), we then obtain the second-order result of the fermion mediated interaction as
\begin{align}
\Gamma_{\rm mi }(i\omega_q, \bq)  =g_{bf}^2\chi_f^{(0)}(i\omega_q,\bq).
\label{Gamma0}
\end{align}
Here $\chi_f^{(0)}(i\omega_q,\bq)$ is the so-called polarization function, or the Lindhard function of a free Fermi gas~\cite{Fetter} and is given by 
\begin{align}
\chi_f^{(0)}(i\omega_q,\bq) &= {T}\sum_{i\omega_k}\int\frac{d^3\bk}{(2\pi)^3} G_f(i\omega_k,\bk) G_f(i\omega_k+i\omega_\bq,\bk+\bq)  \nn\\
& = \int\frac{d^3\bk}{(2\pi)^3} \frac{\theta(k_f-|\bq|)- \theta(k_f-|\bk+\bq|)}{i\omega_q - (\epsilon_{f,\bk+\bq} - \epsilon_{f,\bk})} \nn \\
& = \frac{m_fk_f}{4\pi^2}\left [ -1 +\frac{k_f}{2q}\left (1-\Lambda^2_+ \right )\ln\frac{-1+\Lambda_+}{1+ \Lambda_+} + \frac{k_f}{2q}\left (1-\Lambda^2_-\right ) \ln\frac{-1-\Lambda_-}{1- \Lambda_-}\right ],
\end{align}
where \begin{align}
\Lambda_\pm(i\omega_q, q) = \frac{i\omega_q}{q}\frac{1}{v_f}  \pm \frac{q}{2k_f}
\end{align}
with $v_f = k_f/m_f$ as the Fermi velocity. From Eq.~(11) of the main text we then find to the second order of $a_{bf}$
\begin{align}
a_{\rm eff} = \frac{m_b}{4\pi} g_{bf}^2\lim_{\bq\rightarrow 0} \chi_f^{(0)}(0,\bq) = -\frac{1}{2\pi} \left(\frac{m_f}{m_b} + \frac{m_b}{m_f}+2 \right ) k_f a^2_{bf},
\end{align}
where $k_f$ is given by Eq.~(\ref{kf0}). The perturbative results shown in Fig.~4  of the main text are calculated using this formula. 

Finally, substituting Eq.~(\ref{Gamma0}) into Eq.~(5) and Eqs.(7)-(8) of the main text, we find 
\begin{align}
\label{G110}
G_{11}(i\omega_p,\bp) &= \frac{i\omega_p+ \epsilon_{b,\bp} + n_b g_b +n_b g^2_{bf} \chi_f^{(0)}(i\omega_p,\bp)}{-\omega_p^2 - E^{(0)2}_{b,\bp} - 2n_b g^2_{bf} \epsilon_{b,\bp} \chi^{(0)}_{f}(i\omega_p,\bp)}  \\
G_{12}(i\omega_p,\bp) &=- \frac{n_bg_b+ n_b g^2_{bf} \chi_f^{(0)}(i\omega_p,\bp)}{-\omega_p^2 - E^{(0)2}_{b,\bp} - 2n_b g^2_{bf} \epsilon_{b,\bp} \chi^{(0)}_{f}(i\omega_p,\bp)}
\label{G120}
\end{align}
where $E^{(0)}_{b,\bp} = \sqrt{\epsilon^2_{b,\bp} + 2n_bg_{b} \epsilon_{b,\bp} }$ is the Bogoliubov spectrum of the pure BEC. The retarded Green's functions can be obtained from the above temperature Green's functions by analytic continuation, i.e., $G_{11}(\omega,\bp)  = G_{11}(i\omega_p\rightarrow \omega+i0^+,\bp)$. Equations (\ref{G110}) and (\ref{G120}) can be used to calculate the Bogoliubov excitation spectrum $E_{b,\bp}$ and the dynamic structure factor $S_b(\omega,\bq)$ in the perturbative regime. The former satisfies the following equation
\begin{align}
E^{2}_{b,\bp}    - E^{(0)2}_{b,\bp}  -  2n_b g^2_{bf} \epsilon_{b,\bp} \chi^{(0)}_f(E_{b,\bp}  ,\bp) = 0,
\end{align}
while the latter is defined by   
\begin{align}
  S_b(\omega,\bq) = \frac{1}{\pi}{\rm Im}\chi_b(\omega ,\bp),
  \end{align}
where the retarded density response function $\chi_b(\omega,\bp)$ with real frequency $\omega$ is obtained directly from the imaginary frequency one $\chi_b(i\omega_p,\bp)$ by analytic continuation. Within the Bogoliubov framework, we can write  $\chi_b(\omega,\bp)$  as
\begin{align}
\chi_b(\omega,\bp) = -N_b\left [G_{11}(\omega,\bp)  + G_{11}(-\omega,-\bp)  + 2G_{12}(\omega,\bp)  \right ],
\label{chibsm}
\end{align}
 where $N_b$ is the total number of bosonic atoms. Using Eqs.~(\ref{G110})-(\ref{G120}) in the above expression we then find~\cite{Yip2001SM} 
\begin{align}
\chi_b(\omega,\bp) = -\frac{2N_b\epsilon_{b,\bp}}{\omega^2 - E^{(0)2}_{b,\bp} - 2n_b g^2_{bf} \epsilon_{b,\bp} \chi^{(0)}_{f}(\omega,\bp)}.
\end{align}

\section*{compressibility sum rule}
\label{sumrules}
In this section, we prove that our strong-coupling theory satisfies the compressibility sum rule. Sum rules are exact identities obeyed by various response functions of a system.  Since any theory treating a resonant Bose-Fermi mixture involves approximations that are non-perturbative in nature, enforcing the sum rules is an important way to gauge the validity of these approximations. We show that once the fermion self-energy is specified by Fig.~2(a) of the main text, fulfillment of the compressibility sum rule with respect to the bosonic dynamic structure factor $S_b(\omega,\bq)$ requires that the bosonic self-energies must include the contributions from the mediated interaction given by Fig.~2(c)-(d), in addition to that given by Fig.~2(b).  

The compressibility sum rule reads~\cite{Pitaevskii2016SM}
\begin{align}
\lim_{\bp\rightarrow 0}\int_{0}^\infty d\omega \frac{S_{b}(\omega,\bp)} {\omega} = \frac{N_b}{2m_b c^2_{b,\rm com}}
\label{csr}
\end{align}
 where 
\begin{align}
 c_{b,\rm com} = \sqrt{({n_b}/{m_b})\left.({\pa \mu_b}/{\pa n_b})\right |_{\mu_{f}}}
 \label{csv}
 \end{align}
  is the compressibility sound velocity. 
  
 Now using  the Kramers-Kronig relation and the fact that ${\rm Im}\chi_{b}(\omega,\bp)$ is an odd function of $\omega$, the left-hand side of Eq.~(\ref{csr}) can be written as 
 \begin{align}
  \int_{0}^\infty d\omega \frac{S_{b}(\omega,\bp)} {\omega} =   \frac{1}{2\pi}\int_{-\infty}^\infty d\omega \frac{{\rm Im}\chi_{b}(\omega,\bp)} {\omega}  = \frac{1}{2}\chi_b(0,\bp).
  \label{kk}
  \end{align} 
Using Eq.~(\ref{chibsm}) and Eqs.~(7)-(8) of the main text, we find
\begin{align}
\chi_b(\omega,\bp) = -2N_b \frac{\epsilon_{b,\bp}+\Sigma_S(\omega,\bp)-\Sigma_{12}(\omega,\bp)-\mu_b}{[\omega-\Sigma_A(\omega,\bp)]^2 -[\epsilon_{b,\bp}+\Sigma_S(\omega,\bp) -\mu_b]^2 + \Sigma_{12}^2(\omega,\bp)},
\end{align}
where
\begin{align}
\Sigma_{A}(\omega,\bp) &= \frac{\Sigma_{11}(\omega,\bp) -\Sigma_{11}(- \omega,-\bp)}{2} \nn \\
\Sigma_S(\omega,\bp) &= \frac{\Sigma_{11}(\omega,\bp) + \Sigma_{11}(-\omega,-\bp)}{2}.
\end{align}
Since $\Sigma_A(\omega,\bp)$ by definition is an odd function of $\omega$ and an even function $\bp$, we have $\Sigma_A(0,\bp) = 0$. Thus
\begin{align}
\lim_{\bp\rightarrow 0}\chi_b(0,\bp) &= 2N_b \lim_{\bp\rightarrow 0} \frac{\epsilon_{b,\bp}+\Sigma_S(0,\bp)-\Sigma_{12}( 0 ,\bp)-\mu_b}{[\epsilon_{b,\bp}+\Sigma_S( 0,\bp) -\mu_b]^2 - \Sigma_{12}^2(0,\bp)} \nn \\
& = 2N_b\frac{1}{\Sigma_S( 0,0) +  \Sigma_{12}(0,0)-\mu_b }.
\label{chib0}
\end{align}
From Eq.~(5) of the main text, we have
\begin{align}
\Sigma_S(0,0)& = 2n_bg_b + n_b\sum_k G_f(k)\calT_{bf}(k) + n_b\Gamma_{\rm mi}(0,0;0) ,\\
\Sigma_{12}(0,0) & = n_bg_b + n_b\Gamma_{\rm mi}(0,0;0).
\end{align}
Using these expressions  and Eqs.~(9)-(10)  of the main text we then find
\begin{align}
\Sigma_S( 0,0) +  \Sigma_{12}(0,0)-\mu_b = 2n_b [g_b+ \Gamma_{\rm mi}(0,0;0) ] = 2n_b\left. \frac{\pa \mu_b}{\pa n_b }\right |_{\mu_f}.
\label{Sigmamu}
\end{align}
Combining Eqs.~(\ref{kk}), (\ref{chib0}) and (\ref{Sigmamu}), we arrive at Eq.~(\ref{csr}), i.e.,  the compressibility sum rule.  From the above analysis we see that including the self-energy terms given by Fig.~2(c)-(d) is necessary for the proof. In other words, if we only consider the bosonic self-energy given by Fig.~2(b) we would find that 
\begin{align}
 \Sigma_S( 0,0) +  \Sigma_{12}(0,0)-\mu_b = 2n_bg_b,
 \end{align} 
 which would then violate the compressibility sum rule.

\section{fermion quasi-particle dispersion and density}
\label{fermions}
In this section, we provide more details on the calculation of the fermion quasi-particle dispersion and density in the Bose-Fermi mixture. Although we use the formalism of  temperature Green's functions, we set the temperature $T=0$ at the end of all calculations. 

Now, the Bose-Fermi T-matrix can be determined analytically and is given by~\cite{Fratini2010SM,Fratini2012SM}  
\begin{equation}
\mathcal T_{bf}(i\omega_p,\bp)=\frac{g_{bf}}{1-g_{bf}\Pi_{bf}(i\omega_p,\bp)}
\end{equation}
where the Bose-Fermi pair propagator is
\begin{equation}
 \Pi_{bf}(i \omega_p, p)=\frac{m_b}{4\pi^2 p}\times \left[ \frac{2m_r}{m_b} k_fp+\frac {1}{2}(k_f^2- A^2)\ln\frac{k_f+ A}{k_f- A}-\frac{1}{2}(k_f^2- B^2)\ln\frac{k_f+ B}{k_f- B}  \right].
\end{equation}
Here 
\begin{equation}
\begin{split}
 A(i \omega_p,  p)=\sqrt{2m_r(i \omega_p+k_f^2/2m_f-p^2/2M)}+ \frac{m_r}{m_b}p ;\\
 B(i \omega_p, p)=\sqrt{2m_r(i \omega_p+k_f^2/2m_f -p^2/2M)}- \frac{m_r}{m_b}p,
\end{split}
\end{equation}
where $M = m_b + m_f$, $m_r = m_bm_f/M$ and $k_f \equiv (6\pi^2 n_f)^{1/3}$. Thus it is clear that the T-matrix depends on the fermion density $n_f$. 
The quasi-particle dispersion is determined by the poles of the fermion Green's function
\begin{align}
G_f (i\omega_p,\bp) =\frac{1} {i\omega_p - (p^2/m_f-\mu_{f,res})- n_b \calT_{bf}(i\omega_p,\bp) }.
\label{Gf}
\end{align}
The fermion density $n_f$, under the chemical potential fixed by the reservoir, is determined by the equation
\begin{equation}
n_f=T\sum_{i\omega_p }\int\frac{d^3\bp}{(2\pi)^3} G_f(i\omega_p,\bp)=  T\sum_{i\omega_p }\int\frac{d^3\bp}{(2\pi)^3}\frac{1}{i\omega_p-p^2/2m_f+\mu_{f,res}-n_b\calT_{bf}(i\omega_p,p) }.
\label{nf}
\end{equation}
 Since $ \calT_{bf}(i\omega_p,\bp)$ inside $G_f(i\omega_p,\bp)$ depends on $n_f$, we need to solve for the fermion density $n_f$ self-consistently from Eq.~(\ref{nf}) and in doing so we also determine the poles of the Green's function simultaneously. 

\begin{figure}[th]
	\centering
	\includegraphics[width=8cm]{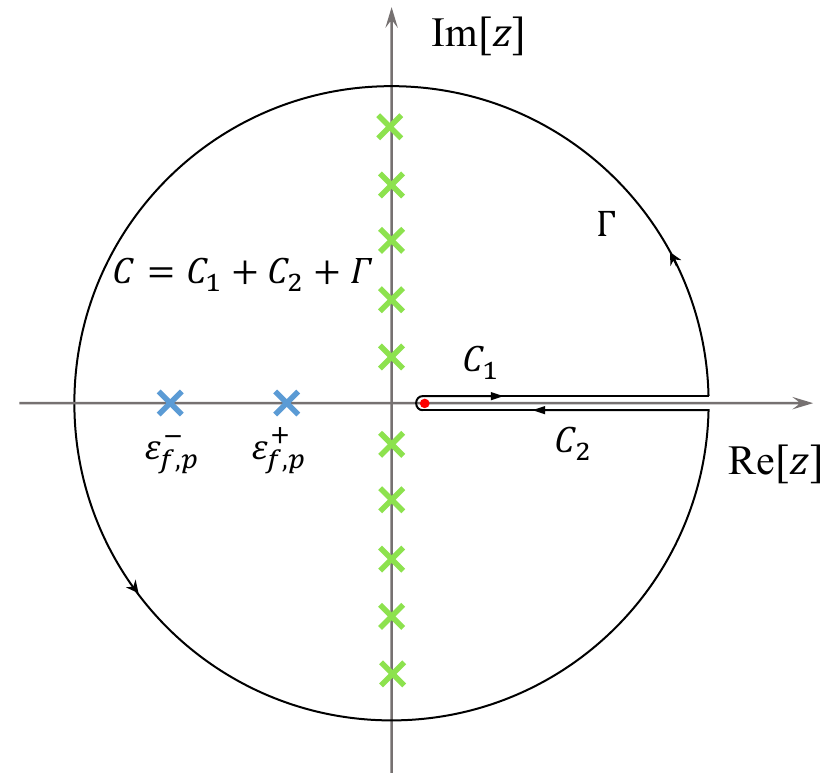}
	\caption{The contour used in performing the Matsubara summations in Eq.~(\ref{nf}) and Eq.~(\ref{aefft}). }
	\label{contour}
\end{figure}

To evaluate the right hand side of Eq.~(\ref{nf}) for a given $n_f$, we first perform the Matsubara summation via a contour integration illustrated by Fig.~\ref{contour}. Since the branch cut introduced by the T-matrix is on the positive real axis, the integration along it vanishes in the $T \rightarrow 0$ limit due to the Fermi-Dirac distribution. We then find
\begin{equation}
n_f=\frac{1}{2\pi^2}\int_0^\infty dp~ p^2  \left[\frac{n_{\rm FD}(\varepsilon_{f,\bp}^+)}{1-n_b\dot {\calT}_{bf} (\varepsilon_{f,\bp}^{+},\bp)} + \frac{n_{\rm FD}(\varepsilon_{f,\bp}^-)}{1-n_b\dot {\calT}_{bf} (\varepsilon_{f,\bp}^{-},\bp)}\right],
\label{nf1}
\end{equation}
where $n_{\rm FD}(x) = \frac{1}{e^{x/T} + 1}$ is the Fermi-Dirac distribution, $\varepsilon_{f,\bp}^{\pm}$ ($\varepsilon_{f,\bp}^{+}>\varepsilon_{f,\bp}^{-}$) are the two poles of the Fermi Green's function satisfying  
\begin{equation}
\label{poleEq}
\varepsilon_{f,\bp}^{\pm}-p^2/2m_f+\mu_{f,res}-n_b{\calT}_{bf}(\varepsilon_{f,\bp}^{\pm},\bp)=0,
\end{equation}
and $\dot\calT_{bf}(\varepsilon,p)$ denotes the derivative with respect to  $\varepsilon$. The two branches of quasi-particle dispersion, $\varepsilon_{f,\bp}^{+}$ and $\varepsilon_{f,\bp}^{-}$, are referred to the repulsive (upper) and attractive (lower) branch respectively. We note that $\varepsilon_{f,\bp}^{\pm}$ so defined  is actually the fermion quasi-particle energy measured from the chemical potential $\mu_{f,res}$. Thus for both branches $\varepsilon_{f,\bp}^{\pm}<0$ with respect to occupied states. In Fig.~3(a) of the main text, we have plotted the quasi-particle energy in absolute terms, namely they correspond to $\varepsilon_{f,\bp}^{\pm} + \mu_{f,res}$ here.

Now, Eq.~(\ref{nf1}) is used if both branches are occupied. However, for $a_{bf} > 0$ the fermions are prepared in the repulsive branch in the weak Bose-Fermi interaction limit. If we assume that an adiabatic tuning of $a_{bf}$ from weak to strong is carried out, the fermions are expected to stay in the repulsive branch. Thus in this case we self-consistently solve 
\begin{equation}
n_f=\frac{1}{2\pi^2}\int_0^\infty dp~ p^2  \frac{n_{\rm FD}(\varepsilon_{f,\bp}^+)}{1-n_b\dot {\calT}_{bf} (\varepsilon_{f,\bp}^{+},\bp)}
\end{equation}
to determine the fermion density and the quasi-particle dispersion. 
Similarly for $a_{bf} < 0$ the fermions are prepared in the attractive branch in the weak Bose-Fermi interaction limit. Assuming that they occupy the attractive branch for all  $a_{bf} < 0$, we then self-consistently solve 
\begin{equation}
n_f=\frac{1}{2\pi^2}\int_0^\infty dp~ p^2   \frac{n_{\rm FD}(\varepsilon_{f,\bp}^-)}{1-n_b\dot {\calT}_{bf} (\varepsilon_{f,\bp}^{-},\bp)}. 
\end{equation} 
The results shown in Fig.~3 of the main text are obtained this way. 

\section{ effective scattering length}
In this section, we provide more details on the calculation of the effective scattering length  $a_{\rm eff}$. From Eq.~(6) and (11) of the main text, we find 
\begin{align}
a_{\rm eff}&=\frac{m_b}{4\pi}T\sum_{i\omega_p}\int \frac{d^3\bp}{(2\pi)^3}\frac{  \calT^2_{bf}(i\omega_p,\bp) }{ \left[ i\omega_p-(p^2/2m_f-\mu_{f,res})-n_b  \calT_{bf}(i\omega_p,\bp) \right]^2 } \\
& = \frac{m_b}{4\pi}T\frac{\pa }{\pa n_b}\left [ \sum_{i\omega_p}\int \frac{d^3\bp}{(2\pi)^3}\frac{  \calT_{bf}(i\omega_p,\bp) }{  i\omega_p-(p^2/2m_f-\mu_{f,res})-n_b  \calT_{bf}(i\omega_p,\bp)  } \right ].
\label{aefft}
\end{align}
Performing the Matsubara summation in a similar way to that in Eq.~(\ref{nf}), we find
\begin{align}
a_{\rm eff}&=\frac{m_b}{8\pi^3} \sum_{\alpha = \pm}\frac{\pa }{\pa n_b} \int_0^\infty dp~ p^2  \frac{\calT_{bf} (\varepsilon_{f,\bp}^{\alpha},\bp)n_{\rm FD}(\varepsilon_{f,\bp}^\alpha)}{1-n_b\dot {\calT}_{bf} (\varepsilon_{f,\bp}^{\alpha},\bp)}.
\label{aeff1}
\end{align}
Since $\varepsilon^\alpha_{f,\bp}$ also depends on $n_b$ through Eq.~(\ref{poleEq}), we find 
\begin{equation}
\frac{\pa }{\pa n_b}\varepsilon^{\pm}_{f,\bp}- \calT_{bf}-n_b\dot \calT_{bf} \frac{\pa }{\pa n_b}\varepsilon^{\pm}_{f,\bp}=0.
\end{equation}
Using this expression in Eq.~(\ref{aeff1}) we finally obtain
 \begin{align}
a_{\rm eff}&=\frac{m_b}{8\pi^3} \sum_{\alpha=\pm} \int_0^\infty dp \,p^2 \left\{   \frac{ 2\calT_{bf}(\varepsilon^\alpha_{f,\bp},\bp) \dot T_{bf }(\varepsilon^\alpha_{f,\bp},\bp) }{ \left[1-n_b\dot \calT_{bf}(\varepsilon^\alpha_{f,\bp},\bp) \right]^2} n_{\rm FD}(\varepsilon^\alpha_{f,\bp}) \right. \nn \\
&\left.+\frac{ n_b\ddot \calT_{bf}(\varepsilon^\alpha_{f,\bp},\bp) T^2_{bf} (\varepsilon^\alpha_{f,\bp},\bp) }{ \left[1-n_b\dot \calT_{bf} (\varepsilon^\alpha_{f,\bp},\bp) \right]^3} n_{\rm FD}(\varepsilon^\alpha_{f,\bp})+ \frac{\calT_{bf}^2(\varepsilon^\alpha_{f,\bp},\bp)}{  \left[1-n_b\dot \calT_{bf} (\varepsilon^\alpha_{f,\bp},\bp) \right]^2}\dot n_{\rm FD} (\varepsilon^\alpha_{f,\bp})  \right\},
\label{aeff2}
 \end{align}
 where $\ddot{\calT}_{bf}(\varepsilon,\bp)$ denotes the second order derivative with respect to $\varepsilon$. Again, for the results shown in Fig.~4 of the main text, we include only the contribution from the repulsive branch  for $a_{bf} > 0$ and only the contribution from the attractive branch for $a_{bf} < 0$.
\section{bosonic dynamic structure factor }
\begin{figure}[h]
	\centering
	\includegraphics[width=7.5cm]{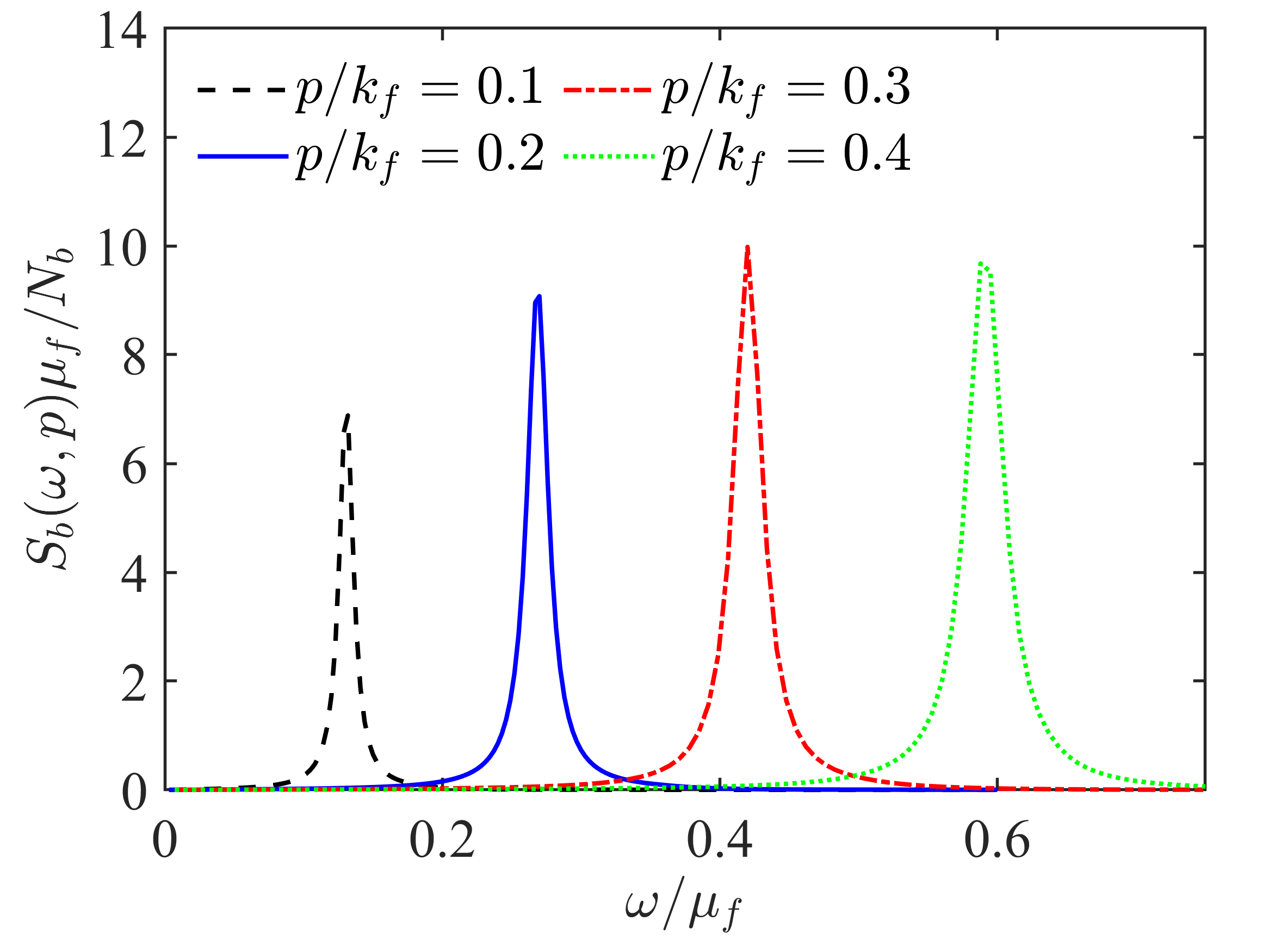}
		\includegraphics[width=7.5cm]{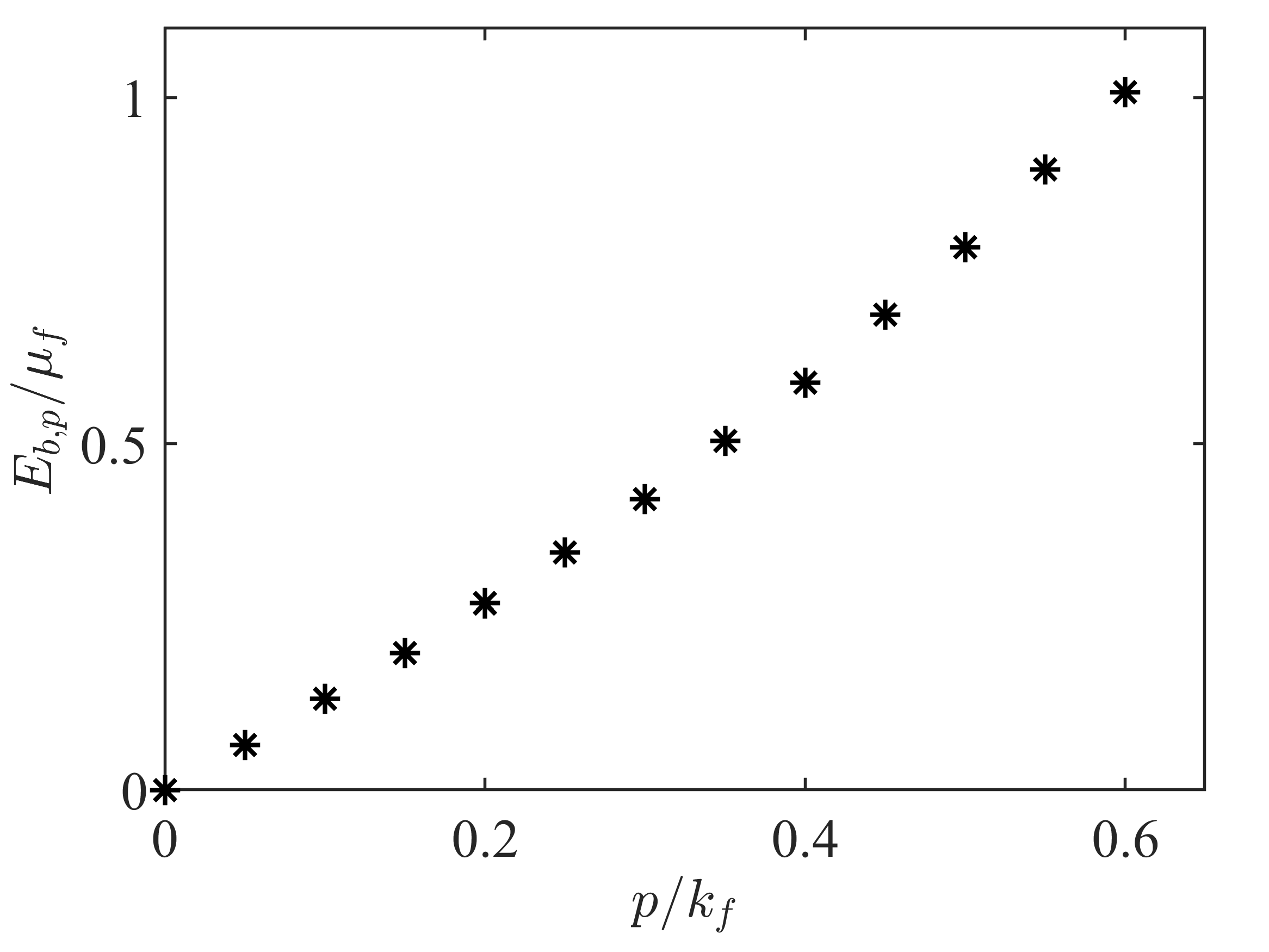}
	\caption{left: the dynamic structure factor; right: the Bogoliubov spectrum. Here $n_b/n_f = 10$, $1/k_fa_{bf} = -15$, $k_fa_b = 0.067$ and  $c^{(0)}_b/v_f = 0.65$. }
	\label{sb2}
\end{figure}

\begin{figure}[h]
	\centering
	\includegraphics[width=7.5cm]{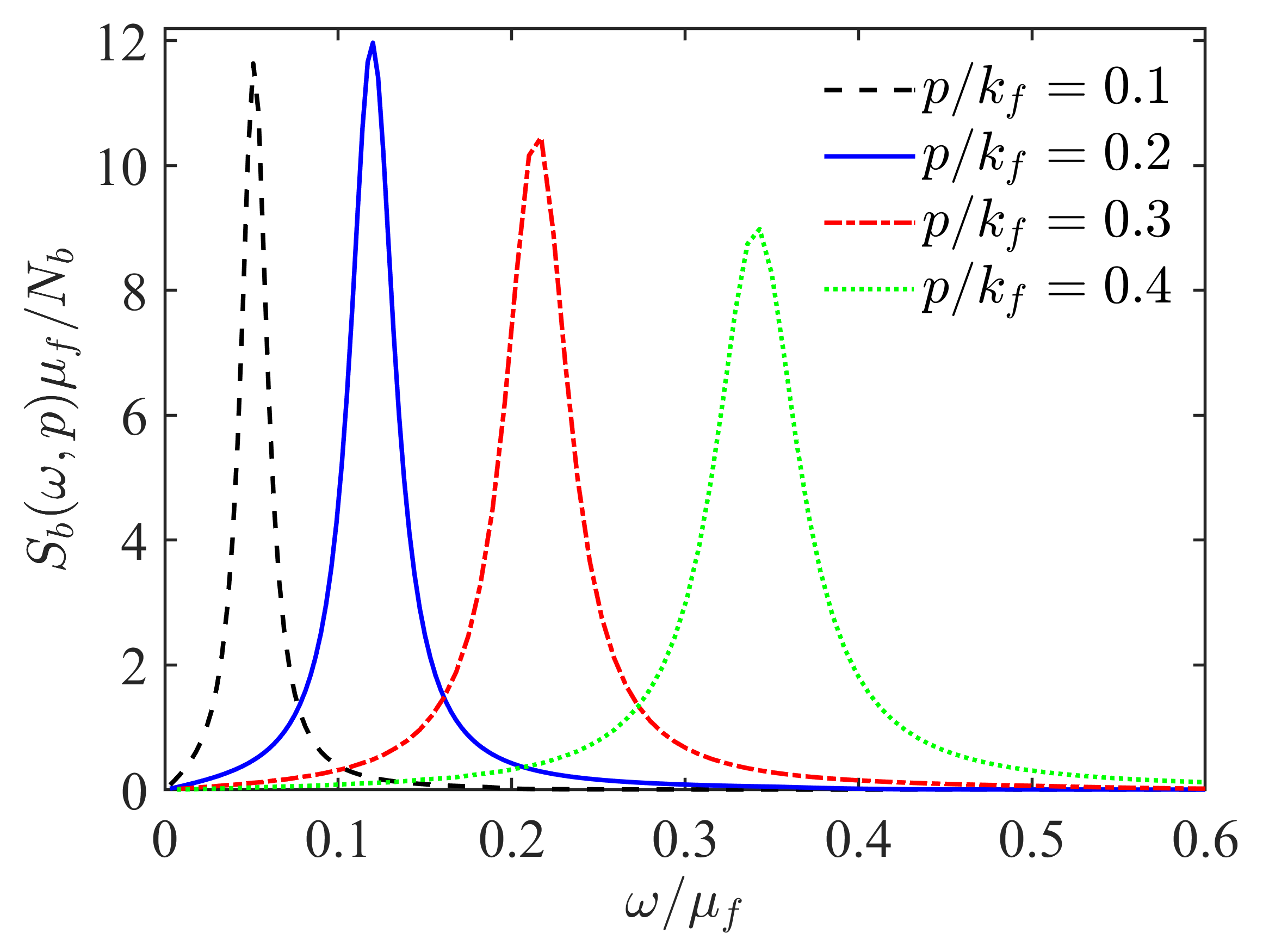}
		\includegraphics[width=7.5cm]{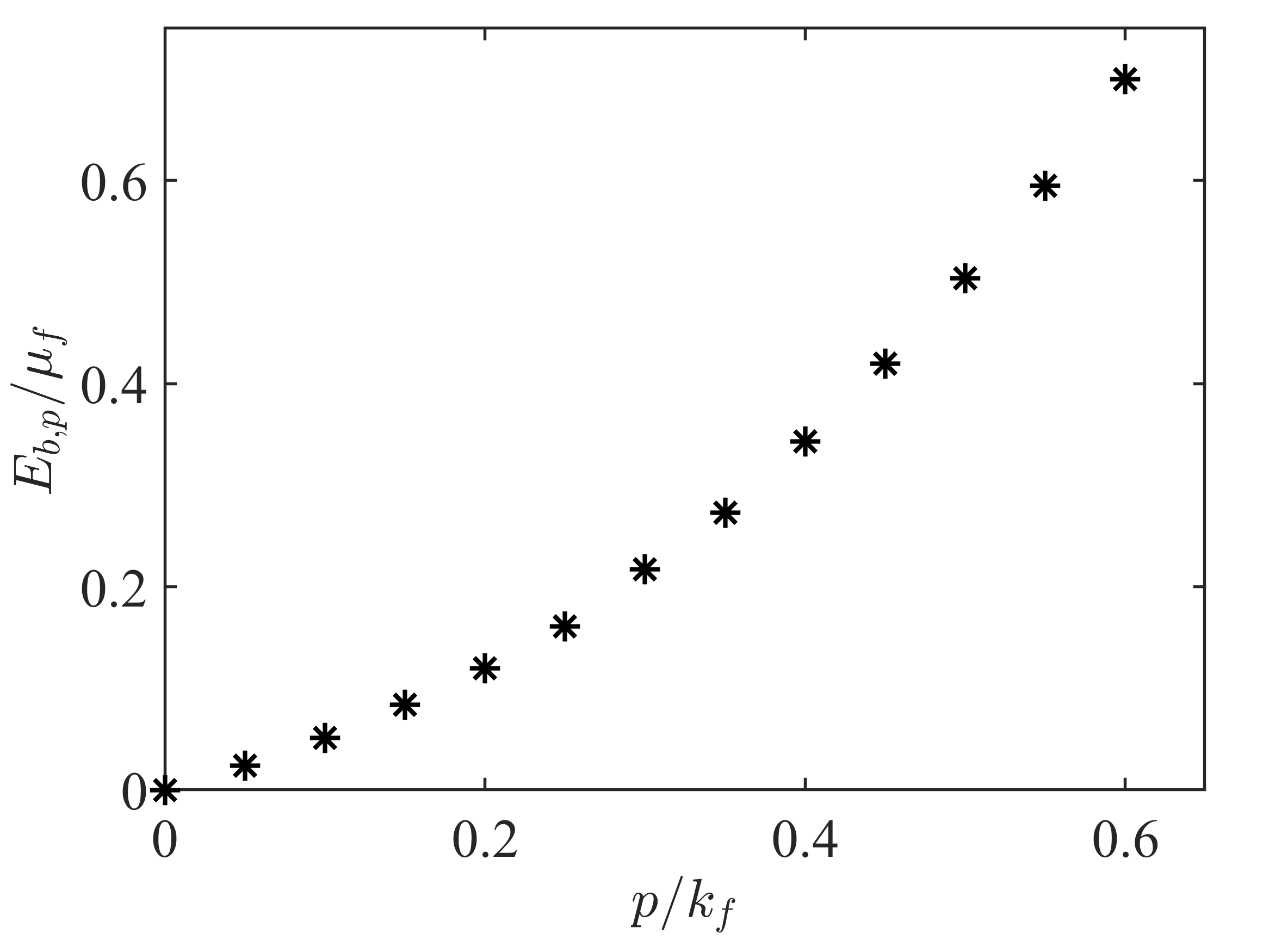}
	\caption{left: the dynamic structure factor; right: the Bogoliubov spectrum. Here $n_b/n_f = 1$, $1/k_fa_{bf} = -3$, $k_fa_b = 0.14$ and  $c^{(0)}_b/v_f = 0.30$.}
	\label{sb1}
\end{figure}

In this section, we provide additional results on the bosonic dynamic structure factor $S_{b}(\omega,\bq)$ for the $^{23}$Na-$^{40}$K mixture. Because the numerical calculations of the self-energies are largely similar to those of the effective scattering length $a_{\rm eff}$, we will not repeat these details here.  Here we include the results of $S_{b}(\omega,\bq)$ for the $^{23}$Na-$^{40}$K mixture at two sets of parameters that are representative of the following two scenarios respectively: (i) the Bose-Fermi scattering length $a_{bf}$ is small but $c_b^{(0)}/v_f $ is comparable to $1$; and (ii)  
$c_b^{(0)}/v_f $ is much smaller than $1$ but $a_{bf}$ is still large.  In the first scenario the Bose-Fermi coupling is weak but the retardation effect is still significant while the opposite is true in the second scenario. The results shown in Fig.~\ref{sb2} correspond to scenario (i) and those in Fig.~\ref{sb1} correspond to scenario (ii).  We can see that in both cases the dynamic structure factor exhibits a single peak, meaning that only the bosonic sound mode has been excited. The underlying reason is that the bosonic sound mode and the induced fermionic zero sound mode are weakly coupled either when the Bose-Fermi coupling is weak or the retardation effect is small.  

\end{document}